\documentclass[sigconf]{acmart}

\DeclareMathOperator{\sign}{sign}

\usepackage{amsmath}
\usepackage{mathtools}
\usepackage{todonotes}
\usepackage{booktabs}
\usepackage{tabularx}
\usepackage{array}
\newcolumntype{L}[1]{>{\raggedright\let\newline\\\arraybackslash\hspace{0pt}}m{#1}}
\newcolumntype{C}[1]{>{\centering\let\newline\\\arraybackslash\hspace{0pt}}m{#1}}
\newcolumntype{R}[1]{>{\raggedleft\let\newline\\\arraybackslash\hspace{0pt}}m{#1}}
\AtBeginDocument{%
  \providecommand\BibTeX{{%
    \normalfont B\kern-0.5em{\scshape i\kern-0.25em b}\kern-0.8em\TeX}}}

\acmConference[AdvML '22]{AdvML '22: Workshop on Adversarial Learning Methods for Machine Learning and Data Mining}{August 15, 2022}{Washington DC, USA}
\acmBooktitle{AdvML '22: Workshop on Adversarial Learning Methods for Machine Learning and Data Mining, August 15, 2022, Washington DC, USA}
\begin{document}

\title{Machine-learned Adversarial Attacks against Fault Prediction Systems in Smart Electrical Grids}

\author{Carmelo Ardito}
\email{carmelo.ardito@poliba.it}
\affiliation{%
  \institution{Polytechnic University of Bari}
  \streetaddress{P.O. Box 1212}
  \city{Bari}
  \country{Italy}
}

\author{Yashar Deldjoo}
\email{yashar.deldjoo@poliba.it}
\affiliation{%
  \institution{Polytechnic University of Bari}
  \city{Bari}
  \country{Italy}}

\author{Tommaso~Di Noia}
\email{tommaso.dinoia@poliba.it}
\affiliation{%
  \institution{Polytechnic University of Bari}
  \city{Bari}
  \country{Italy}
}

\author{Eugenio~Di Sciascio}
\email{eugenio.disciascio@poliba.it}
\affiliation{%
 \institution{Polytechnic University of Bari}
 \city{Bari}
 \country{Italy}}

\author{Fatemeh Nazary}
\authornote{Corresponding author: Fatemeh Nazary (fatemeh.nazary@poliba.it). Authors are listed in alphabetical order.}
\email{fatemeh.nazary@poliba.it}
\affiliation{%
  \institution{Polytechnic University of Bari}
  \city{Bari}
  \country{Italy}}
  
  \author{Giovanni Servedio}
\email{g.servedio@studenti.poliba.it}
\affiliation{%
  \institution{Polytechnic University of Bari}
  \city{Bari}
  \country{Italy}}




\renewcommand{\shortauthors}{C. Ardito, Y. Deldjoo, T. Di Noia, E. Di Sciascio, F. Nazary, G. Servedio}

\begin{abstract}
In smart electrical grids, fault detection tasks 
may have a high impact on society due to their economic and critical implications. In the recent years, numerous smart grid applications, such as defect detection and load forecasting, have embraced data-driven methodologies. 
The purpose of this study is to investigate the challenges associated with the security of machine learning (ML) applications in the smart grid scenario. 
Indeed, the robustness and security of these data-driven algorithms have not been extensively studied in relation to all power grid applications.  We demonstrate first that the deep neural network method used in the smart grid is susceptible to adversarial perturbation. Then, we highlight how studies on fault localization and type classification illustrate the weaknesses of present ML algorithms in smart grids to various adversarial attacks.
\end{abstract}

\maketitle

\section{Introduction and Context}




Over the years, conventionally-operated electrical grids have undergone major revisions and upgrades in terms of dependability, robustness, and efficiency, thus moving to what we call today Smart Grids (SG). One of the most critical components of SGs is their application in fault detection, fault classification, and routine examination of the underlying disruptions that trigger the failures. Power grid networks are inherently vulnerable to physical damages, and electrical faults can be caused by natural accidents such as tree falling on a power line, a bird contact, lightning or aging of the equipment~\cite{DBLP:journals/swevo/SantisRS18}. 
Power grid faults may lead to large-scale cascading effects, which might have a devastating impact on the economy and security of a country. As a result, rapid detection and classification of faults with a high degree of fidelity is a key service for the Electric Power Supply industry and the overall security of the critical energy infrastructure (CEI)~\cite{onyeji2014cyber}.

This paper is focused towards classification of faults and their occurring area. Fault zone classification (FZC) aims to find the zone (or the exact location) in which the fault has occurred, while in fault type classification (FCT) the main objective is to determine the fault type class. Voltage sags are the main cause of faults, which can manifest as asymmetric phase-to-phase (LL), single-phase-to-ground (LG), or two-phase-to-ground (LLG) or symmetric three-phase-to-ground (LLLG or LLL) faults in both transmission and distribution systems~\cite{DBLP:journals/tsg/ShiZMD19,DBLP:journals/tsg/AbdelgayedMS18}. Previous literature has utilized a combination of tools and techniques from electrical engineering, signal processing, and artificial intelligence (AI)~\cite{DBLP:journals/tsg/ShiZMD19,DBLP:journals/tsg/DasAS19,sapountzoglou2020generalizable} to solve the above fault classification tasks. Among them, machine-learned (ML) models, notably those based on deep learning, have witnessed an increase in their acceptance in the current infrastructure of power systems, owing to the huge amounts of data spanning energy networks.



\begin{figure}[bhtp]
    \centering
    \includegraphics[width = 0.85\linewidth]{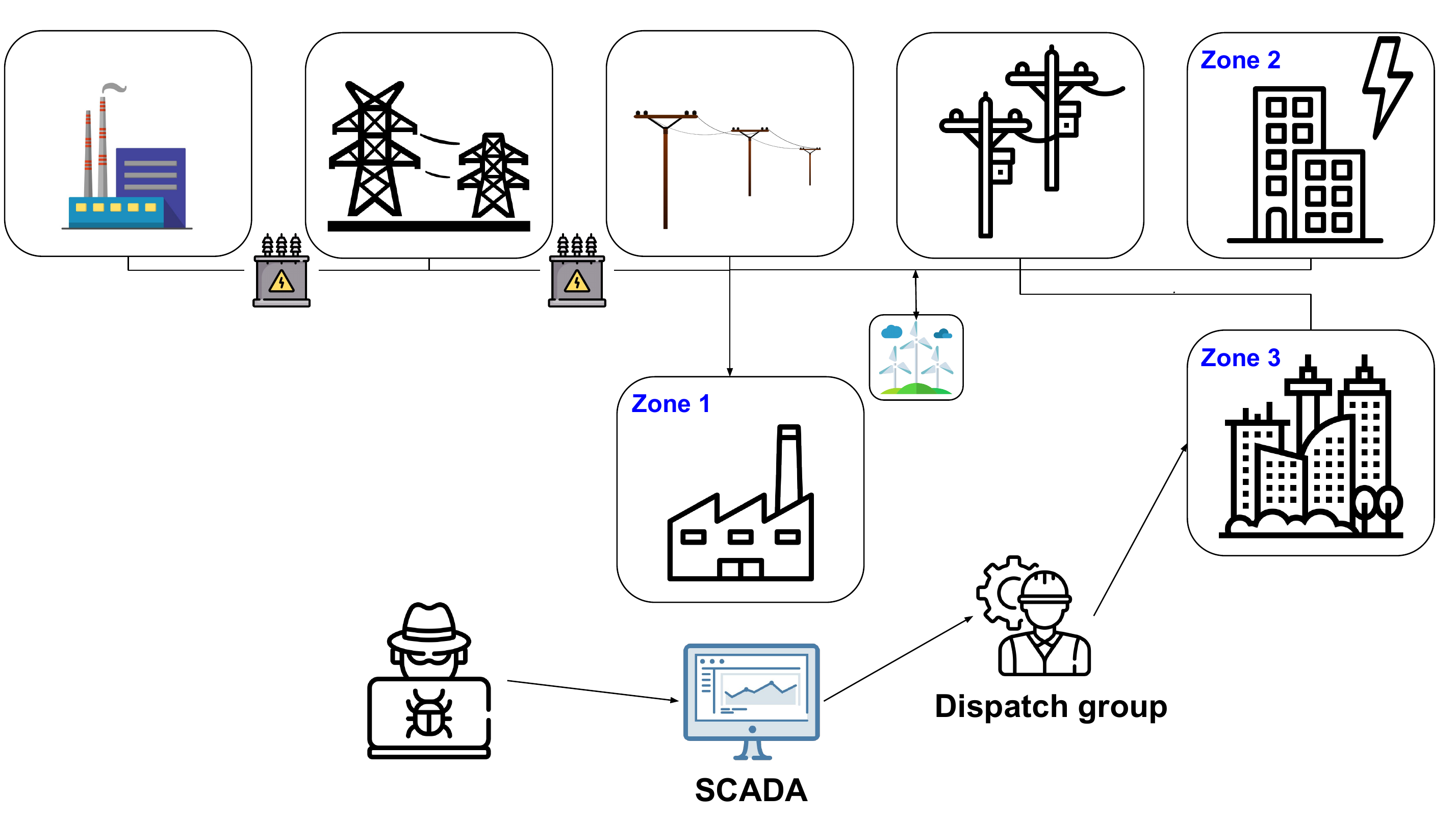}
    \caption{A hypothetical illustration of targeted adversarial attacks against fault zone prediction in smart grids. As a result of an adversarial attack on a fault location prediction system, dispatch recovery groups were dispatched to zone 3 by accident rather than zone 2, where they belonged.}
    \label{fig:overview}
\end{figure}

Unfortunately, notwithstanding their great performance, the intricacy of current (deep) inference methods may be their downfall. Adversarial attacks can take advantage of their vulnerabilities to compromise the confidentiality, integrity or availability of SGs (aka the CIA triad)~\cite{zarreh2019risk,DBLP:journals/tifs/DoMC16}. Adversarial attacks are operationalized via \textit{adversarial examples} -- subtle but non-random  perturbations -- designed to induce an ML model to produce erroneous outputs (e.g., to misclassify an input sample). As seen in Figure~\ref{fig:overview}, an attacker can enter the SG system's communication network in order to attack the failure prediction system used in supervisory control and data acquisition (SCADA)~\cite{DBLP:journals/jnca/CuiQGXY20}. The purpose of the targeted adversarial attack is to induce the SCADA fault classification system's machine learning (ML) model to misclassify an input sample as belonging to a known but incorrect class. Toward this aim, in the FZC scenario, the attacker selects an illegitimate target class label to prolong the rescue operation. By providing false positive signals, they can misdirect the rescue squad to places that do not need help.

\noindent\textbf{Adversarial examples and attack on SG}.
\citet{DBLP:conf/smartgridcomm/ChenTD18} discuss the vulnerability of machine learning algorithms used in building load forecasting and power quality disturbance analysis against specific adversarial attack. \citet{DBLP:journals/tsg/Farajzadeh-Zanjani21} developed a generative-adversarial system for partially labeled samples named semi-supervised generative adversarial learning. 
Adversarial attacks on convolutional neural network-based event causes have been presented in \cite{DBLP:conf/isgt/NiazazariL20} for three specific power grid events: (1) line energization, (2) capacitor bank energization, and (3) fault prediction. The fast gradient sign technique (FGSM) \cite{DBLP:journals/corr/SzegedyZSBEGF13} was used to introduce minor perturbations into voltage or current data for adversarial crafting. Additionally, the Jacobian-based Saliency Map Attack (JSMA) is used to compare the level of the FGSM's adversary. Finally, adversarial training improves the CNN classifier's performance against specific attacks. According to \cite{DBLP:conf/eenergy/SongTR021}, voltage stability is assessed using adversarial instances generated using techniques such as FGSM, PGD, DeepFool, and Universal Adversarial Network (UAN), as well as Universal Adversarial Perturbation (UAP). Adversarial training is used to protect against these adversarial examples.

Our key contributions are summarized as follows:
\begin{itemize}
    \item We investigate the impact of adversarial attacks against several key fault classification problems, and their combination on a widely used dataset based on the IEEE-13 test node feeder;
    \item We analyze adversarial attacks by examining multiple experimental situations with different adversarial goals;
    \item We show, via empirical experiments on a dataset collected from the IEEE-13 test node feeder, that adversarial attacks can degrade the quality of classifications significantly.
\end{itemize}

\section{Approach}
\label{sec:4 approach}
We have conducted adversarial attacks against two machine-learned fault classification task in smart electrical grids, which serve as the core attack target. 

\noindent\textbf{Adversarial task.} Given a training dataset $\mathcal{D}$ of $n$ pairs $(x, y) \in \mathcal{X} \times \mathcal{Y}$, where $x$ is the input sample, and $y$ is its corresponding class label, the classification problem is formulated as finding a target function $f_\theta: \mathcal{X} \rightarrow \mathcal{Y}$ that can predict the class label $y$ surroundings the input sample $x$, where $\theta$ is the model parameter. The goal of the adversarial attacks is to find a non-random perturbation $\delta$ to produce an adversarial example $x^{adv} = x + \delta$ such that it can induce an inaccurate detection (e.g., mis-classification). The methods by which $delta$ is learned are referred to as \textit{adversarial attacks}, and they can be either targeted or untargeted \cite{anelli2021study,deldjoo2019assessing}.

\begin{definition}[Targeted adversarial attack]\label{def:tar_attack}
Given a trained classifier $f(\textbf{x}; \theta)$ and a test instance from the dataset $\textbf{x}_0 \in \mathcal{D}$ where $f(\textbf{x}_0; \theta)= y_0$, the goal of a targeted attack is to perturb $\textbf{x}_0$ with a small budget $\left\lVert \delta \right\rVert \leq \epsilon$ such that the perturbed sample would be mis-classified to the target label $y_T \neq y_0$, referred to as the \textit{mis-classification label}. The problem can be represented using an unconstrained optimization problem formulation
    \begin{equation}
        \label{eq:att_trg2}
        \begin{aligned}
            & \underset{\delta: \left\lVert \delta \right\rVert \leq \epsilon }{\text{min}}
            & & \mathcal{L}(f(\textbf{x}_0+\delta; \theta), \ y_T) 
        \end{aligned}
    \end{equation}  
\noindent One can note that in this case, here the attacker aims to \textbf{minimize} the distance (loss) between the adversarial prediction $f(\textbf{x}_0+\delta)$ and the mis-classification label $y_T$.
\end{definition}

\begin{definition}[Untargeted  attack] The goal of the attacker in untargeted attack is to cause any mis-classification to \textit{maximize} the loss between the adversarial prediction and the legitimate label $y_0$
\begin{equation}
\label{eq:att_untrg_2}
\begin{aligned}
    & \underset{\delta: \left\lVert \delta \right\rVert \leq \epsilon }{\text{max}}
    & &\mathcal{L}(f(\textbf{x}_0+\delta; \theta), \ y'\neq y_0) 
\end{aligned}
\end{equation} 
as such, it is clear that the attacker's objective in this scenario is to cause any mis-classification $y'$, regardless the of the specific type.
\end{definition}

\noindent\textbf{Fault Classification in Smart Grids.}\label{sec:4.1}
We consider different multi-class classification problems pertinent to fault prediction in smart grids with $K \geq 2$ classes in this paper, in which $\mathcal{X}$ is
the input space and $\mathcal{Y} = \{1, 2, . . . , K\}$ the output space. Our problem showcases two different target labels for the problems at hand: (i) fault location and (ii) fault type. Therefore, the main task is split into three sub-tasks:
\begin{enumerate}
    \item Fault location classification (FLC): with $K=4$ the task aims to classify a given signal into its originating zone as shown in Table \ref{tab:dataset}.
    \item Fault type classification (FTC): with $K=11$  the task aims to classify a given signal into one of predefined fault types as shown in Table \ref{tab:dataset}.
    \item Joint location and type classification (FLC+FTC) $k=44$ integrating the both fault class labels in the preceding cases; 
\end{enumerate}
where, (1) and (2) are explicitly contained in the dataset, while (3) is derived by combing each different possible combination of task (1) and task (2). Thus, we can state that the joint task is expected to be a more complex one compared to the former.


\noindent\textit{Adversary goal.} The adversary is interested in mis-classifying smart-grid fault classification tasks in each of the three FZC, FTC, and joint sub-tasks through the use of two types of attacks: untargeted vs. targeted. In the latter situation, the purpose may be to produce more difficult-to-reach or difficult-to-resolve (mis-classification) labels in order to obstruct or delay the recovery task.\\
\noindent\textit{Adversary knowledge.} Our assumption is a \textit{white-box} setting where the attacker has full access to the feature extraction model parameters and input features that would be altered due to attacks. In a targeted attack scenario, the attacker can also obtain the class labels. \\

\section{Experimental Evaluation}\label{sec:5}
We studied attacks on smart grids using data from the IEEE-13 test node feeder and explain the experimental setup below.



\begin{table}[!t]
\centering
\caption{The characteristic of the dataset used for training the machine-learned fault classification models in this work.}
\footnotesize
\begin{tabular}{cll}
\hline
Item & \multicolumn{2}{c}{Details} \\ \hline
Fault type & \multicolumn{2}{l}{\begin{tabular}[c]{@{}l@{}}phase to ground        AG, BG, CG\\ phase to phase          AB, AC, BC \\ phase to phase to ground         ABG, ACG, BCG\\ three phase               ABC \\three phase to ground              ABCG \end{tabular}} \\ \hline
Fault location & \multicolumn{2}{l}{\begin{tabular}[c]{@{}l@{}}zone 1                        branch 632-671\\ zone 2                        branch 632-633\\ zone 3                        branch 692-675\\ zone 4                        branch 671-680\end{tabular}} \\ \hline
Fault resistance & \multicolumn{2}{l}{\begin{tabular}[c]{@{}l@{}}0.0010,                       0.0273,
0.0535,  0.0798\\                        0.1061, 
0.1323  0.1586,  0.1848\\                               0.2111, 0.2374,  0.2636, 0.2899\\   0.3162,   0.3424, 0.3687, 0.3949\\
  0.4212, 0.4475, 0.4737, 0.5, 1, 2\\
\end{tabular}} \\\hline
\end{tabular}
\normalsize
\label{tab:dataset}
\end{table}
\subsection{Datasets}
For data collection and creating the training dataset for the fault classification in smart grids,
similar to~\cite{DBLP:journals/access/ShafiullahA18,DBLP:journals/tsg/AbdelgayedMS18,DBLP:journals/tsg/ShiZMD19} 
we used short-circuit faults that were injected to IEEE-13 node test feeder using the MATLAB Simulink environment. The node feeder contained renewable energies such as wind turbine and photovoltaic system.
We divided the network into four zones, adjacent to four load flow buses (numbered via  671,633, 675, and 680, see~\cite{onaolapo2019simulation}), and measured the three-phase voltage signals. 


We applied 11 short circuit faults to four specified zone in the IEEE-13 network. These faults cover every conceivable short-circuit 
faults and are summarized in Table~\ref{tab:dataset}. To ensure having a sufficient number of samples in the training dataset, each fault was generated with 22 different fault resistance values~\cite{DBLP:journals/access/ShafiullahA18,DBLP:journals/tsg/HossanC19}.
Our final training dataset contained $4~\text{(zones)} \times 11~ \text{(faults)} \times 22~\text{(resistance values)} \times 4~\text{(measured locations)} = 3872$ samples.
Note that we collected (measured) signals from 4 locations regardless of locations, and after feature extraction (see below) stacked them together to create a super-vector which was fed into the neural network ML model. 

To inject faults, the entire simulation duration was carried out in the time interval $t= [0.0 - 0.022]$, with the network frequency $60 Hz$, sampling time $0.00001$. Each fault with every resistance was applied at a certain start time $t= 0.01$ and revoked at a specified end time $t= 0.02$, hence $t_f= [0.01-0.02]$ represents the faulty duration and $t_h= [0-0.01]$ represents the healthy duration. For the signal type, in this work we only relied on (three-phase) voltage signals and kept investigation of other possible signals such as current for future investigation.







\begin{figure*}
    \centering
    \includegraphics[width=0.75\linewidth]{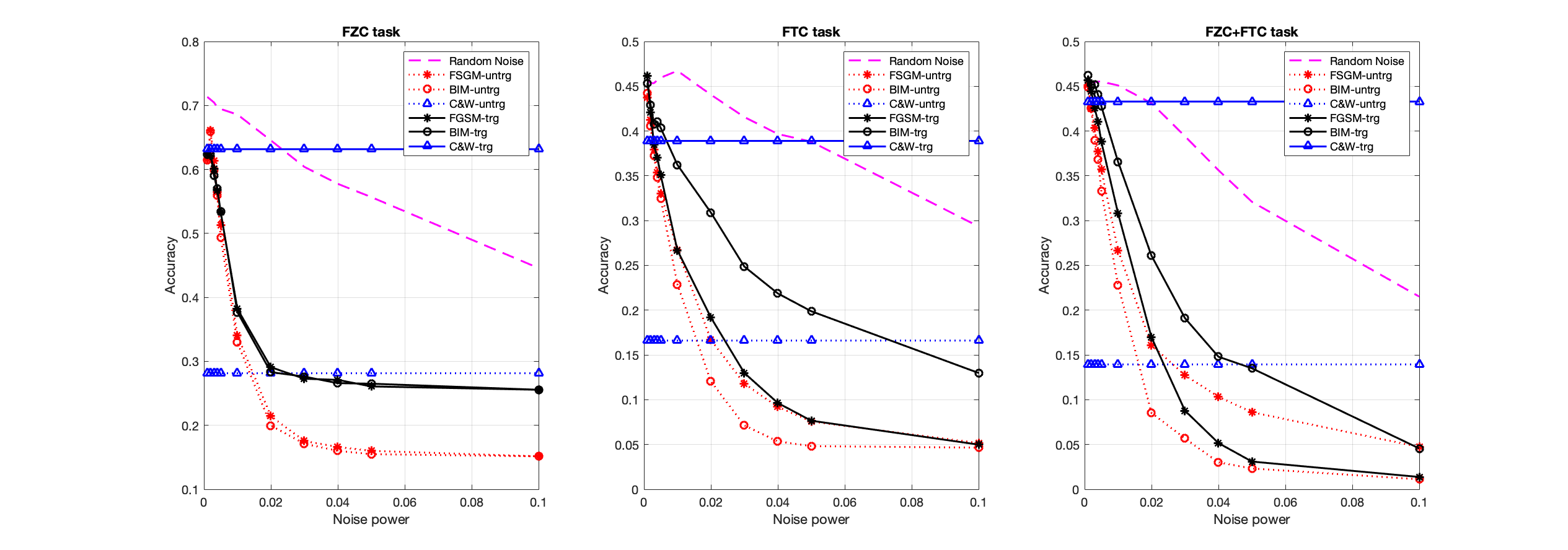}
    \caption{Three tasks under targeted and untargeted adversarial attacks. Classification accuracy for $FZC = 0.7134$, $FTC = 0.4569$, and $FZC + FTC = 0.4543$. Best results for C\&W were obtained under $\ell_{\infty}$ for untargeted attacks and $\ell_{2}$ for targeted attacks. Note that the starting point of noise power for all attacks and random noise is $0.001$.} 
    \label{fig:evaluation}
\end{figure*}

The time series signals were represented as discrete features retrieved from the time, frequency, and wavelet domains using temporal, Discrete Fourier transform (DFT), and Discrete wavelet transform (DWT) analysis, as previously explored~\cite{DBLP:conf/itasec/ArditoDSN21,DBLP:journals/tsg/SalehHE17}. Afterwards, we extract from each domain, six features related to energy, maximum, as well as the $4$-th moment of their probability distribution functions (PDFs) (e.g., mean, norm, skewness, kurtosis). The overall length of the feature vectors utilized in the learning model is 48, divided into $6~\text{(time)} + 6~\text{(DFT)} + 36~\text{(DWT)}$, where we employed 6 (coefficients) ~$\times$~6 (aggregation operations) for the DWT features, resulting in a 36-dimensional feature vector.





\subsection{Adversarial Attacks} 
The performed attacks consist of the fast gradient sign method (FGSM), basic iterative method (BIM) \cite{DBLP:conf/iclr/KurakinGB17a}, and Carlini and
Wagner (C\&W) \cite{DBLP:journals/corr/CarliniW16}, see also \citet{deldjoo2021survey}.
FGSM is a white-box attack that employs the sign of the loss function's gradient to learn adversarial perturbations and BIM is the iterative version of the FGSM. Formally, in the untargeted scenario, FGSM aims to generate a perturbation that maximizes the training loss formulated as
    \begin{equation}
        \label{eq:att_linf}
        \delta = \epsilon \cdot \sign    (\bigtriangledown_x \ell(f(x;\theta),y))
    \end{equation}
where $\epsilon$ (perturbation level) represents the attack strength and $\bigtriangledown_x$ is the gradient of the loss function w.r.t. input sample $\textbf{x}$, $y$ is the legitimate label and $\sign(.)$ is the sign operator. A targeted FGSM attack is, instead, formulated as
    \begin{equation}
        \label{eq:att_linf_targeted}
        \delta = - \epsilon \cdot \sign    (\bigtriangledown_x \ell(f(x;\theta),y_T))
    \end{equation}
in which the goal of the attacker is maximize the conditional probability $p(y_T|x)$ for a given input $x$.


 The second category of adversarial attacks is Carlini and Wagner. It is a powerful attack model for finding adversarial perturbation under three various distance metrics ($\ell_0$, $\ell_2$, $\ell_{\infty}$). Its key insight is similar to L-BFGS \cite{DBLP:journals/corr/SzegedyZSBEGF13} as it transforms the constrained optimization problem into an empirically chosen loss function to form an unconstrained optimization problem as
\begin{equation}
    \underset{\delta}{\text{min}} \ \left( \left\lVert \delta \right\rVert_p^p + c \cdot h(\textbf{x}+\delta, y_T) \right)
\end{equation}
where $h(\cdot)$ is the candidate loss function.\qed

The C\&W attack has been used with several norm-type constraints on perturbation $\ell_0$, $\ell_2$, $\ell_{\infty}$ among which the $\ell_2$ and $\ell_{\infty}$-bound constraint has been reported to be most effective~\cite{DBLP:journals/corr/CarliniW16}.

\section{Experiments and Results}
\subsection{Explored Machine-Learnings Tasks}

\textbf{Model and training details.} We trained a deep neural network, a Multi-layer Perceptron (MLP), for the three classification tasks specified in Section 2. The model is made of an input layer, two dense layers, and an output layer. The latter is the only layer that varies throughout the three tasks, as its number of neurons must correspond to the number of output classes in each task. The tasks require separate training phases, which all take place with the same settings, using 500 Epochs, Adam Optimizer, and fixed learning rate of 10e-3 with a batch-size of 20. The hyper-parameters were obtained after fine-tuning.

\noindent \textbf{Implementation of the attacks.} We employed the IBM Adversarial Robustness Toolbox to perform the adversarial attacks due to its full compatibility with Keras and its wide offer of suitable attacks for a deep learning model. The performed attacks consist of FGSM, multi-step (BIM), and C\&W attacks. These attacks were performed in both untargeted and targeted scenarios.

\subsection{Results}\label{sec:6}
\textbf{Evaluation Questions.} To obtain a better understanding of the effectiveness of the examined adversarial attacks against fault classification system in SGs, through the course of experiments, we intend to answer the following evaluation questions.

\begin{description}
    \item \textbf{RQ 1}: Against the three faults classification tasks in SGs presented in Section~\ref{sec:4.1}, how effective are adversarial perturbations generated by different adversarial attack methods (FGSM, BIM, and C\&W) compared to random noise? 
    \item \textbf{RQ 2}: How does the performance of attacks change when we alternate between the ~\textbf{attack targets}?
\end{description}
\vspace{-2.5mm}


\noindent \textbf{Discussion.} We begin our experimental study by addressing the above evaluation questions.

\textit{Answer to RQ~1.} This research question verifies whether the
application of adversarial attacks against fault classification system (FZC, FTC, and joint) has a sensible impact on the behavior of the ML models. As shown in Figure~\ref{fig:evaluation}, all investigated adversarial attacks FGSM, BIM, and C\&W have a much greater impact than random perturbation across three tasks and under different noise levels ($\epsilon$), with the effect growing as the perturbation budget increases. Comparing the strength of the three adversarial attack models, BIM is the strongest in all tasks. For instance, in the case of (untargeted, FTC) with attack budget (noise level) equal to $\epsilon = 0.04$, BIM untargeted adversarial attack accuracy reaches $0.05$, whilst FGSM and C\&W reach $0.09$ and $0.16$, respectively, under the same condition. The effect of attack target (targeted vs. untargeted) is stronger on BIM and C\&W than on FSGM. For example, for the FTC ($\epsilon = 0.04$), the classification accuracy is 0.21 vs. 0.05 (BIM-untargeted vs. BIM-targeted), while for FGSM the corresponding difference is only 0.1 vs. 0.09 (FGSM-untargeted vs. FGSM-targeted). 

\textit{In summary, the attacks' powers might be contrasted according to $\textbf{BIM>C\&W>FGSM}$ (the first being the strongest). The lone exception is \textbf{C\&W-targeted}, which deviates from the trend and performs poorly, while \textbf{C\&W-untargeted} performs well in all the explored scenarios.}

\textit{Answer to RQ~2.} This research question verifies how much the performance of difference adversarial attacks varies across smart grid fault prediction tasks, and whether complexity of these tasks impacts the performances obtained.

We start this by assessing the absolute power of attacks across three tasks. At $\epsilon = 0.04$ the power of attacks FGSM-untrg, BIM-untargeted, C\&W-untargeted, FGSM-targeted, BIM-targeted, C\&W-targeted is equal to 0.166, 0.160, 0.281, 0.271, 0.265, and 0.631 respectively. Thus, w.r.t the base ML model (0.713), we may remark a relative degradation of 329\% , 345\% , 153\% , 163\%, 168\%, and 13\%. The equivalent relative degrading power of attacks for FTC task are 396\%, 756\%, 175\%, 374\%, 108\%, 17\% and for the joint FZC+FTC task include 339\%, 1408\%, 226\%, 779\%, 206\%, 4.9\%. Thus, the average degradation power for (untargeted, targeted) goals are, FZC=(275.6\%, 114.6\%), FTC=(442.3\%, 166.3\%), FTC=(657.6\%, 329.9\%). We might notice that both untargeted and targeted attack models work better (are stronger) as the task gets more complicated and this is true for both types of tasks.

\textit{In summary, the result of empirical evaluation shows that the complexity of the fault prediction tasks (in SGs) impacts the effectiveness of the explored adversarial attacks, meaning the attacks are better able to manipulate the decision outcomes according to \textbf{FZC+FTC$>$FTC$>$FZC}.}

\vspace{-2mm}

\section{Conclusion}\label{sec:7}
This work examines the security of fault classification systems in smart electrical grids powered by deep neural networks. Minor adversarial perturbations can reduce the quality of fault classification systems, highlighting the need for further studies to defend against adversarial training and detection methods (see~\cite{deldjoo2021survey}). Visual explanation of such adversarial threats~\cite{ardito2022visual} would constitute another interesting direction, which future work will investigate. Additionally, multi-party computation techniques, such as federated learning, could be used to develop privacy-preserving fault-prediction systems~\cite{anelli2019towards}, allowing separate zones to train models without exchanging data with a central server. Some of these challenges mirror issues in recommender systems \cite{anelli2021put,anelli2021study,deldjoo2019assessing,deldjoo2018content,DBLP:journals/ia/AnelliDNF20,anelli2021msap} that we have addressed in our previous work, and we aim to apply similar strategies in the context of machine learning.


\begin{acks}
This work has been partially funded by \textit{e-distribuzione S.p.A} company, Italy, through a PhD scholarship granted to Fatemeh Nazary.
\end{acks}



\bibliographystyle{ACM-Reference-Format}
\bibliography{refs}

\end{document}